\documentclass[draft]{agujournal}
\draftfalse

\journalname{Geophysical Research Letters}

\begin{document}

%
%

\title{On production of gamma rays and Relativistic Runaway Electron Avalanches from Martian dust storms}

%
%

 \authors{Shahab Arabshahi\affil{1},
 Walid A. Majid\affil{1}, Joseph R. Dwyer\affil{2}, and
 Hamid K. Rassoul\affil{3}}

\affiliation{1}{Jet Propulsion Laboratory, California Institute of Technology, Pasadena, California, USA.}
\affiliation{2}{Department of Physics, University of New Hampshire, Durham, New Hampshire, USA.}
\affiliation{3}{Department of Physics and Space Sciences, Florida Institute of Technology, Melbourne, Florida, USA.}


\correspondingauthor{Shahab Arabshahi}{shahab.arabshahi@jpl.nasa.gov}

\begin{keypoints}
\item Simulated the production of energetic electron avalanches from Martian dust storms.\ 
\item Found the electron avalanche characteristic length for different electric fields inside Martian dust storms. \ 
\item The gamma rays can be detected by ground-based instruments and satellites orbiting Mars.\ 
\end{keypoints}

%
%
\copyright \ 2017. All Rights Reserved.

\begin{abstract}
Production of runaway electron avalanches and gamma rays originating inside Martian dust storms are studied using Monte Carlo simulations. In the absence of in situ measurements, we use theoretical predictions of electric fields inside dust storms. Electrons are produced through the relativistic runaway electron avalanches process, and energetic photons are results of the bremsstrahlung scattering of the electrons with the air. Characteristic lengths of the runaway electron avalanche for different electric fields and the energy spectrum of electrons are derived and compared to their terrestrial counterparts. It is found that it is possible for Martian dust storms to develop energetic electron avalanches and produce large fluxes of gamma ray photons similar to terrestrial gamma ray flashes from Earth's thunderstorms. The phenomenon could be called Martian gamma ray flash, and due to the very thin atmosphere on Mars, it can be observed by both ground-based instruments or satellites orbiting the planet.
\end{abstract}
%
%

\section{Introduction}
\label{sec:intro}
Occurrence of dust storms on Mars has been known for many decades \citep{Antoniadi1930La1659} and are commonly observed by satellites and telescopes. The storms vary in size from tens of kilometers to planet-encircling storms. The smallest dusty phenomena are convective vortices called dust devils. They usually last no more than about 10~min, have diameters less than 1~km and are no more than 10~km in height \citep{Renno2000MartianData}. The regional dust storms are tens of kilometers in diameter, few kilometers tall, and last 10--20~days. Larger storms are formed when one or several small ones merge and expand. They could then cover large portions of the planet, reaching up to 40~km tall, and lasting from several weeks to few months. The planet-encircling storms are called "global" storms and normally occur in the Martian southern spring and summer \citep{Martin1993AnMars,Cantor2001MartianObservations,Cantor2007MOCStorm}.

Regional Martian dust storms are similar in size to terrestrial dust storms, both having cores much hotter than their surface temperature, and having complicated vortex winds \citep{Ryan1983PossibleMars}. It has also been suggested that Martian dust storms are electrically active \citep{Sentman1991ElectrostaticEnvironment,Melnik1998ElectrostaticStorms,Renno2004MATADORDevils,Renno2012COMMENTSMARS}. However, unlike terrestrial thunderstorms' inductive charging between large graupel and smaller ice and water droplets, electrification at Martian dust storms is the result of triboelectric charging, i.e., friction of dust particles with themselves and the ground. Although there has been some reports on possible observation of electric discharges \citep{Ruf2009EmissionStorm}, the existence of electrical activity on Mars has not been completely confirmed. Theoretical modeling and observations on Earth have shown that it is possible for the storms to develop electric fields 5--25~kV/m \citep{Renno2003ElectricalStorms,Renno2004MATADORDevils,Farrell2006IntegrationDevil}. This surpasses the breakdown electric field, 20~kV/m, in the Martian atmosphere, which has a much lower density than Earth \citep{Melnik1998ElectrostaticStorms}. Such electric fields could result in glow and filamentary optical discharges in the Martian atmosphere \citep{Farrell1999DetectingStorms}.

Such predicted electric fields inside Martian dust storms may also result in production of X-rays and gamma rays similar to those that frequently occur inside terrestrial thunderstorms. On Earth, electrons can run away in electric fields of thunderclouds and gain energies of up to several tens of MeV \citep{Dwyer2008a}. They can then produce beams of X-rays and gamma rays through bremsstrahlung scattering with air molecules and atoms. These beams are so powerful that can travel hundreds of kilometers up in the Earth's atmosphere and are detected by satellites in orbit. 

In this paper, we have investigated the possibility of the production of runaway electrons through the relativistic runaway electron avalanche (RREA) process for the case of Martian atmosphere. We first review the relevant physics of RREA mechanism in Section \ref{sec:rrea summary}. Details of the Monte Carlo simulation of RREA are presented in Section \ref{sec:ream summary}. In Section \ref{sec:results and discussion} we present results of the simulation followed by a discussion of the implications of the results. Finally, we conclude our study in Section \ref{sec:conclusion}.

\section{Relativistic Runaway Electron Avalanches}\label{sec:rrea summary}
\begin{figure*}[t]
 \includegraphics[width=0.75\columnwidth]{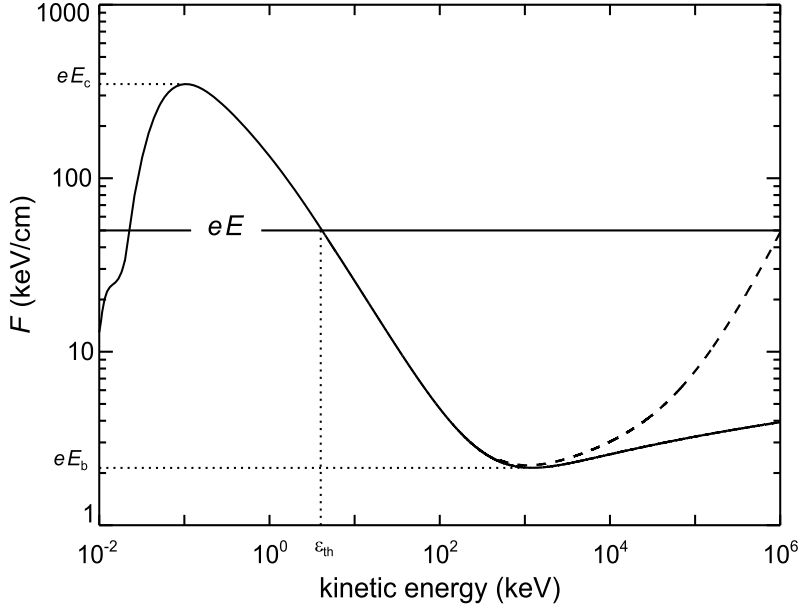}
 \caption[Electron frictional force]{Energy loss and energy gain of electrons or positrons, in air at STP condition, per unit length as a function kinetic energy \citep{InternationalCommissiononRadiationUnitsandMeasurements1984}. Solid curve is the frictional force due to inelastic scattering with air atoms, dashed curve is the frictional force including bremsstrahlung, and the horizontal line is the electric force from a 5000 kV/m electric field $E$.}\label{fig:electron friction}
\end{figure*}

Free electrons would "run away" in a medium where they gain more energy from a background electric field than they lose energy through different interactions and scattering processes. The required electric field depends on the initial kinetic energy of electrons and, for air, it is always greater than $E_{\text{th}} = 2.84 \times 10^{5}$~V/m~$\times~N$ \citep{Dwyer2003}, where $N$ is the ratio of the air density in the medium to its value at sea level. $E_{\text{th}}$ takes into account the energy loss due to elastic scattering. For electrons moving directly along the field lines and without elastic scattering, the minimum electric field is smaller and is called the break-even field $E_{b}=2.18 \times 10^{5}$~V/m~$\times~N$. Figure \ref{fig:electron friction} shows the effective frictional force and the electric force (horizontal line) on an electron (or positron) as a function of its kinetic energy in air. For an electron (or positron) to run away in a region with constant electric field $E$, its initial kinetic energy should be greater than $\varepsilon_{\text{th}}$. At electric fields larger than $E_{c} \approx 3 \times 10^{7}$~V/m~$\times~N$, any free electron, independent of its initial kinetic energy, would run away. \cite{Wilson1925} showed that electric fields inside thunderstorms are large enough to produce runaway electrons. The energetic "seed" electrons can come from cosmic ray air showers or radioactive decays.

Furthermore, it is possible for a single-seed electron with energy above $\varepsilon_{\text{th}}$ and inside a high-enough-field region, to produce an avalanche of new runaway electrons through M{\o}ller scattering with air atoms and molecules \citep{Gurevich1992RunawayThunderstorm}. This is called the relativistic runaway electron avalanche (RREA) process. Moreover, if the region is large enough, multiple avalanches can be developed from the X-rays and positrons that were produced during the runaway process and traveled to the beginning of the region. This can become a self-sustaining discharge called the relativistic feedback discharge (RFD). The runaway electrons can then produce large fluxes of gamma ray photons at the end of the high-field region \citep{Dwyer2003a, Dwyer2007RelativisticAtmospheres}.

RREA can also occur at other planetary atmospheres \citep{Roussel-Dupre2008PhysicalAtmospheres}. Its occurrence at Jupiter \citep{Dwyer2007RelativisticAtmospheres} and at Venus \citep{Bagheri2016AnVenus} has already been investigated and shown to be possible. For Mars, even for the case of near-breakdown fields inside dust storms, it is not clear if RREA would initiate in such a low-density atmosphere. Here we have used the runaway electron avalanche model (REAM) \citep{Dwyer2003a, Dwyer2007RelativisticAtmospheres} to investigate the phenomena at Mars.

\section{Simulation Details}\label{sec:ream summary}
REAM is a three-dimensional Monte Carlo code that simulates the interactions and propagation of electrons, positrons, and photons in any gaseous medium (such as planetary atmospheres) in the presence of electrostatic fields. Included interactions for electrons and positrons are as follows : ionization and atomic excitation (dynamical friction), bremsstrahlung, M{\o}ller (Bhabha) scattering, elastic scattering with a shielded Coulomb potential, direct electron-positron pair production ("trident process") \citep{Vodopiyanov2015TheRates}, and positron annihilation. For photons, photoelectric absorption, Compton scattering, and pair production are modeled. REAM also has some limitations when involving complex electric field configurations and ground effects. The simulation was updated in order to study the characteristics of RREA originating from the Martian atmosphere. The atmosphere of Mars has density of about 0.020 kg/m$^3$ close to its surface. It is mainly composed of carbon dioxide (95.32\%) and nitrogen (2.7\%). Other minor components are argon, oxygen, carbon monoxide, water, nitrogen monoxide, neon, hydrogen-deuterium-oxygen (HDO), krypton, and xenon \citep{Williams2014MarsSheet}. For our simulation we have only considered the main components CO$_2$ and N$_2$ since the small amounts of other species' interaction with energetic electrons and photons do not have significant effect on RREA.

\begin{figure*}[t]
 \includegraphics[width=0.5\columnwidth]{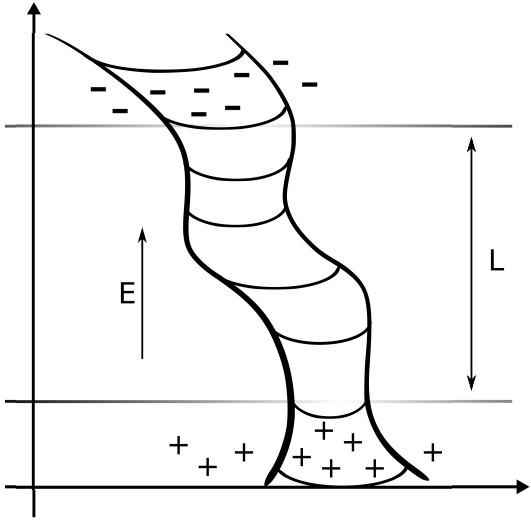}
 \caption{Martian dust storms charge structure and electric field configuration used in REAM.}
 \label{fig:dust devils charge structure}
\end{figure*}

The simulation domain in REAM is carried out in three-dimensional space. The electric field, $E$, is set upward along the vertical component ($z$~axis). 
Laboratory experiments have shown that during triboelectric charging, lighter grains typically become negatively charged and heavier grains become positively charged \citep{Ette1971TheParameters,Lacks2007EffectSystems}. In a simple electrification model, charged vertical winds stratify the grains and create vertically upward electric field as illustrated in Figure \ref{fig:dust devils charge structure}. We have used similar electric field configuration in our simulations.

The vertical extent, $L$, was chosen large enough for the simulation to reach steady state. Although there is no limit in the horizontal extent, most of the particles are produced inside the cone, with base radius equal to $L$. Simulations were performed for the case of Martian atmosphere at near surface. The parameters are summarized in Table \ref{tab:parameters}. The avalanche $e$-folding length and energy spectrum of runaway electrons are obtained and presented in the following section.

\begin{table}
\caption{Summary of atmospheric parameters used in REAM.}
\centering
\begin{tabular}{l c}
\hline
 Parameter  & Value  \\
\hline
  Atmosphere density \citep{Williams2014MarsSheet}  & 0.020~kg/m$^{3}$  \\
  Atmosphere scale height \citep{Williams2014MarsSheet} & 11.1~km \\
  Atmosphere composition  & 95.32\% CO$_{2}$, 2.7\% N$_{2}$   \\
  Seed electron beam energy  & 1 MeV   \\
  Average ionization energy  & $1.4179244 \times 10^{-17}$~J   \\
\hline
\end{tabular}
\label{tab:parameters}
\end{table}

\section{Results and Discussion}\label{sec:results and discussion}
Simulations were performed for different electric fields ranging from 5~kV/m to 50~kV/m. This covers the range of predicted electric fields inside Martian dust storms \citep{Farrell2006IntegrationDevil,Melnik1998ElectrostaticStorms,Renno2004MATADORDevils,Kok2009ElectrificationChemistry}. The total potential difference used in the model, varied from 50~MV to 250~MV. These are greater than the 50 MV minimum potential required for the feedback mechanism to be effective \citep{Dwyer2012}. Although these are large potentials, they are reasonable values for the available potential difference based on our current understanding of the Martian dust storms. There has not been a direct measurement of the storms' electrical potential yet; however, the values should be within the range used in our simulations. This is based on current predictions of their electric fields (5--25 kV/m) and observations of their vertical extent (up to 40 km).

\begin{figure*}[t]
 \includegraphics[width=\columnwidth]{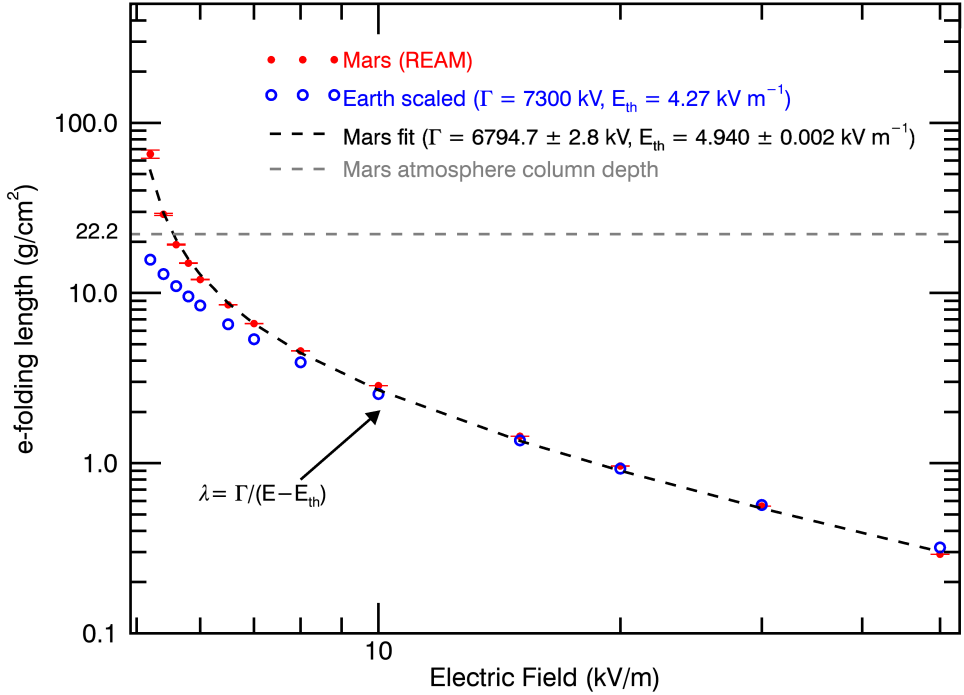}
 \caption{RREA characteristic lengths for electric fields between 5~kV/m to 50~kV/m. The red points show the avalanche characteristic lengths inside Martian atmosphere, blue circles are the lengths when the atomic number density of the Earth's atmosphere is scaled to the Martian value \citep{Dwyer2004ImplicationsLightning}, black dashed curve is the empirical equation~\eqref{eq:lambda} fit to the Martian RREA characteristic lengths (red points), and gray horizontal dashed line shows the total atmospheric depth of Mars.}
 \label{fig:e-folding length}
\end{figure*}

For each tried electric field, we injected a monoenergetic beam of 1 MeV seed electrons into the avalanche region. Such initial seed electrons can be provided by cosmic rays to initiate the first avalanche. Feedback processes would then be the main source of seed electrons and would make the avalanche self-sustaining. Electrons then propagated through all the processes mentioned above, and particles were collected at five different altitudes (hit planes) inside the region. The $e$-folding (characteristic) lengths $\lambda$ were found from comparing the population of electrons at various altitudes inside the high-field region. Figure~\ref{fig:e-folding length} shows the $e$-folding lengths (red circles) of runaway electron avalanches in Martian atmosphere for different electric field values. The black dashed curve is a fit to our results based on the relation proposed by \cite{Dwyer2003a}:
\begin{linenomath*}\label{eq:lambda}
\begin{equation}
\lambda = \frac{\Gamma}{(E-E_{\text{th}}\times N)}
\end{equation}
\end{linenomath*}
$E_{\text{th}}$ and $\Gamma$ are found to be $4.940 \pm 0.002$~kV m$^{-1}$ and $6795 \pm 3 \ kV$. $\Gamma$ denotes the average kinetic energy of runaway electrons per unit charge of an electron ($\overline{K}/e$) and, as previously mentioned, N is the ratio of the air density in the medium to its value at sea level. Blue circles in the figure show the characteristic lengths of RREA when the Earth's empirical relation is scaled to Mars' atmospheric density.

The $e$-folding lengths in Figure \ref{fig:e-folding length} are presented in the atmospheric column depth unit (g/cm$^2$). The figure indicates that for electric fields larger than $\approx$5.5~kV/m, the total column depth of the Martian atmosphere, $\approx$22.2 g/cm$^2$ indicated by the gray dashed line, is large enough to initiate RREA process.

The threshold electric field for the initiation of RREA on Mars, $E_{\text{th}}$, was found to be different, ~14\% larger, than the terrestrial value at similar atmospheric density. This is because the chemical components, i.e., cross section of interactions, are different between the two atmospheres. The average ionization energy in the Martian atmosphere is 88.5~eV, which is larger than the 85.7~eV average ionization energy of air. Although the Bethe-Block relation between ionization energy and dynamical friction is complex, qualitatively speaking, for similar atmospheric densities, larger ionization energies of the gaseous medium results in larger dynamical friction on electrons, which then results in larger electric field threshold required for the electrons to run away.

\begin{figure*}[t]
 \includegraphics[width=\columnwidth]{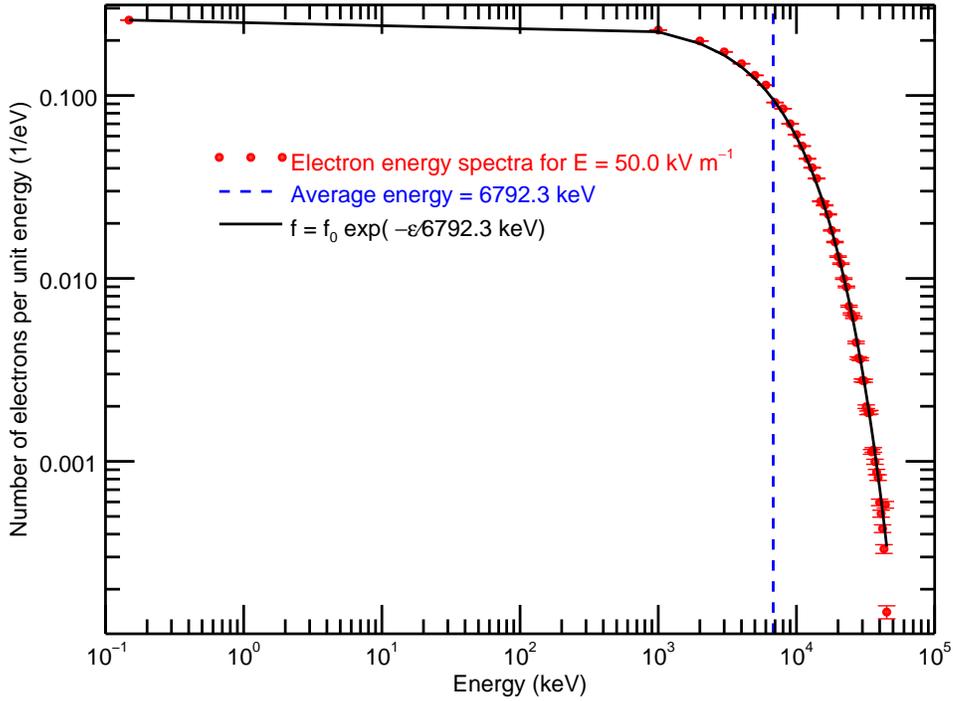}
 \caption{Energy spectrum of electrons produced by RREA mechanism on Mars. Red points represent the energy spectrum of electrons found from REAM, solid black curve shows RREA's analytical spectrum for energies above 100~keV based on equation~\eqref{eq:spectrum}, and the blue dashed line shows the average energy of runaway electrons.}
 \label{fig:spectra}
\end{figure*}

Figure \ref{fig:spectra} shows the energy spectrum of runaway electrons calculated for 50~kV~m$^{-1}$ electric field. The spectrum is independent of the density of the atmosphere and background electric field when $E\gg E_{\text{th}}$, and is described in RREA as an exponential function:
\begin{linenomath*}
\begin{equation}\label{eq:spectrum}
f = f_0 \exp (-\varepsilon/\varepsilon_{\text{avg}})
\end{equation}
\end{linenomath*}

 $\varepsilon_{\text{avg}}$ is the average energy and characteristic energy of electrons and depends on the cross section of the interactions in the medium. On Earth, the average energy $\varepsilon_{\text{avg}}$ varies slightly with the background electric field and is about 7.3~MeV \citep{Dwyer2008Lightning:Applications}. For the case of Mars, we found it to be about 6.7~MeV. In Figure \ref{fig:spectra} the blue dashed line shows the average energy of runaway electrons, and the black solid curve shows RREA's exponential spectrum based on equation~\eqref{eq:spectrum}.
 
Again, the slight difference in the average energies of RREA at the two planets is the result of difference in cross section of interactions that electrons have to undergo. At lower energies, dynamical friction is the dominant interaction. Larger friction value for Mars increases the population of electrons with lower energies. At higher energies, electrons mainly lose energy due to bremsstrahlung scattering. The cross section of bremsstrahlung scattering is proportional to the averaged squared atomic number ($Z^2$) of the gaseous medium. The averaged squared atomic numbers of the Martian atmosphere and air are 54.46 $C^{2}$ and 53.45 $C^{2}$, respectively. Since the two values are comparable, this would result in more or less similar distribution of relativistic electrons. Larger population of low-energy electrons and similar number of runaway electrons make average energy of all electrons smaller for the Martian RREA.

In the absence of any lightning-like discharge or gamma ray observation from Mars, it is difficult to estimate the fluence of gamma ray photons produced by RREA. This is because the total number of photons depends on the characteristics of the storm's electrical environment such as electric potential and storm's charging timescale. However, we know RREA can produce photons with energies from 100 keV to 20 MeV \citep{Dwyer2005AObservations}. Small skin depth of Martian atmosphere for gamma ray photons makes it possible for them to be measured at large distances away from their source.

\section{Conclusion}\label{sec:conclusion}
Using previously predicted electric fields for Martian dust storms, we showed that it is possible for the storms to produce large numbers of energetic electrons from the relativistic runaway electron avalanches (RREAs). Simulations also showed that the avalanche regions inside dust storms could be large enough for RREA to become self-sustaining through feedback processes. This is called the relativistic feedback discharge (RFD) and is a key component in development of terrestrial gamma ray flashes (TGFs). This makes it possible for similar bursts of gamma rays to be produced at Mars as well. Gamma ray photons are the result of the bremsstrahlung scattering of relativistic runaway electrons. For smaller avalanche regions we could have longer but weaker gamma ray emissions similar to the gamma ray glows from thunderstorms. On Earth, TGFs and gamma ray glows commonly occur inside thunderstorms. 

For RREA at Mars, the characteristic length of the avalanche and energy spectrum of electrons are different than terrestrial counterparts. This is due to different chemical components of the atmosphere and interaction cross sections. Direction of the electron and gamma ray photons from RREA at Mars is also different from TGFs. The upward electric field of dust storms produces downward electron avalanche and gamma ray photons. This is in contrast with the mainly upward direction of TGFs. Such downward flux of gamma rays makes their detection more challenging from orbit; however, it also makes them potentially hazardous for future human explorations of the planet. The average energy of the gamma ray photons will be several MeV, which due to the low total atmospheric depth of Mars ($\approx$22.2 g/cm$^{2}$), is capable of traveling long distances and possibly observable by both ground-based instruments or satellites orbiting the planet. The phenomena could be called Martian gamma-ray flashes or MGFs.

%
%
%

\acknowledgments
This work was performed at the Jet Propulsion Laboratory, California Institute of Technology. U.S. Government support is acknowledged. Shahab Arabshahi's research was supported by an appointment to the NASA Postdoctoral Program at the NASA Jet Propulsion Laboratory, California Institute of Technology. The program is administered by Universities Space Research Association under contract with NASA. Requests for data used to generate, or be displayed in figures, graphs, plots, or tables, may be made to the corresponding author (shahab.arabshahi@jpl.nasa.gov).

\end{document}